\documentclass[a4paper,10pt,twoside]{cpc-hepnp}

\usepackage{multicol}
\usepackage{graphicx}
\usepackage{booktabs}
\usepackage{amssymb,bm,mathrsfs,bbm,amscd}
\usepackage[tbtags]{amsmath}
\usepackage{lastpage}

\begin{document}
%\begin{CJK*}{GBK}{song}

\fancyhead[c]{\small Submitted to 'Chinese Physics C'} \fancyfoot[C]{\small \thepage}

\footnotetext[0]{To be submitted}

\title{Single Event Effect Hardness for the Front-end ASICs Applied in BGO Calorimeter of DAMPE Satellite\thanks{supported by the Strategic Priority Research Program on Space Science of the Chinese Academy of Sciences (Grant No.XDA04040202-4) and the Fundamental Research Funds for the Central Universities (Grant No.WK2030040048) }}

\author{%
      GAO Shan-Shan$^{1,2;1)}$\email{ssgao@ustc.edu.cn}%
\quad JIANG Di$^{1,2}$
\quad FENG Chang-Qing$^{1,2;2)}$\email{fengcq@ustc.edu.cn (corresponding author)}%
\quad XI Kai$^{3}$ \\
\quad LIU Shu-Bin$^{1,2}$
\quad AN Qi$^{1,2}$
}
\maketitle

\address{%
$^1$ State Key Laboratory of Particle Detection and Electronics, University of Science and Technology of China, Hefei 230026, China\\
$^2$ Department of Modern Physics, University of Science and Technology of China, Hefei 230026, China \\
$^3$ Institute of Modern Physics, Chinese Academy of Sciences, Lanzhou 730000, China\\
}

\begin{abstract}
Dark Matter Particle Explorer (DAMPE) is a Chinese scientific satellite designed for cosmic ray study with a primary scientific goal of indirect detection of dark matter particles. As a crucial sub-detector, BGO calorimeter measures the energy spectrum of cosmic rays in the energy range from 5 GeV to 10 TeV. In order to implement high-density front-end electronics (FEE) with the ability to measure 1848 signals from 616 photomultiplier tubes on the strictly constrained satellite platform, two kinds of 32-channel front-end ASICs, VA160 and VATA160, are customized. However, a space mission period of more than 3 years makes single event effect (SEE) a probable threat to reliability. In order to evaluate the SEE sensitivity of the chips and verify the effectiveness of mitigation methods, a series of laser-induced and heavy ion-induced SEE tests were performed. Benefiting from the single event latch-up (SEL) protection circuit for power supply, the triple module redundancy (TMR) technology for the configuration registers and optimized sequential design for data acquisition process, VA160 and VATA160 with the quantity of 54 and 32 respectively have been applied in the flight model of BGO calorimeter with radiation hardness assurance.
\end{abstract}



\footnotetext[0]{\hspace*{-3mm}\raisebox{0.3ex}{$\scriptstyle\copyright$}2015
Chinese Physical Society and the Institute of High Energy Physics
of the Chinese Academy of Sciences and the Institute
of Modern Physics of the Chinese Academy of Sciences and IOP Publishing Ltd}%

\begin{multicols}{2}

\section{Introduction}

DAMPE (Dark Matter Particle Explorer), one of the four satellites planned in the framework of the Strategic Pioneer Research Program in Space Science of the Chinese Academy of Sciences, is going to be launched at the end of 2015. Its main scientific objective is to measure cosmic rays with much wider energy range (5 Gev--10 Tev) than existing space experiments in order to identify possible dark matter signatures \cite{lab1}.

As a key sub-detector of DAMPE, BGO (Bismuth Germanate) electromagnetic calorimeter provides a wide range of energy deposition of the particles traversing the detector. In order to reconstruct the electromagnetic shower profile, 308 BGO crystal bars with dimensions of 2.5$\times$2.5$\times$60 cm$^3$ form 14 layers. Crystal bars in consecutive layers are oriented orthogonally to each other. Each crystal bar optically couples two Hamamatsu R5610 photomultiplier tubes (PMTs). The 2nd, 5th and 8th dynodes of each PMT are readout synchronously to achieve a total dynamic range of 2$\times$10$^{5}$ \cite{lab2,lab3}. Hence, to measure 1848 signals on the power-limited and weight-limited satellite platform, two readout chips for PMTs (VA160 and VATA160), are customized to implement high-density and low-power front-end electronics (FEE) \cite{lab4}.

However, single event effect (SEE) is a vital factor that affects the reliability of FEE \cite{lab5}. DAMPE orbits the earth at an attitude of 500 kilometers and an inclination of 97 degrees for the mission period of at least 3 years. The radiation sources of the orbit are composed of galactic cosmic rays, solar particles, and particles trapped in the Van Allen belts, which results in a wide variety of particles with various amounts of kinetic energy corresponding to a wide spectrum of linear energy transfer (LET). Through CREME96, a tool for SEE rate prediction, the LET spectrum of the orbit is calculated, as shown in Fig.~\ref{fig1} \cite{lab6,lab7}. It indicates that the SEE tolerance of semiconductor should be greater than 37.5 MeV$\cdot$cm$^2$/mg, otherwise mitigation methods should be seriously considered to assure radiation hardness.

\begin{center}
\includegraphics[width=8cm]{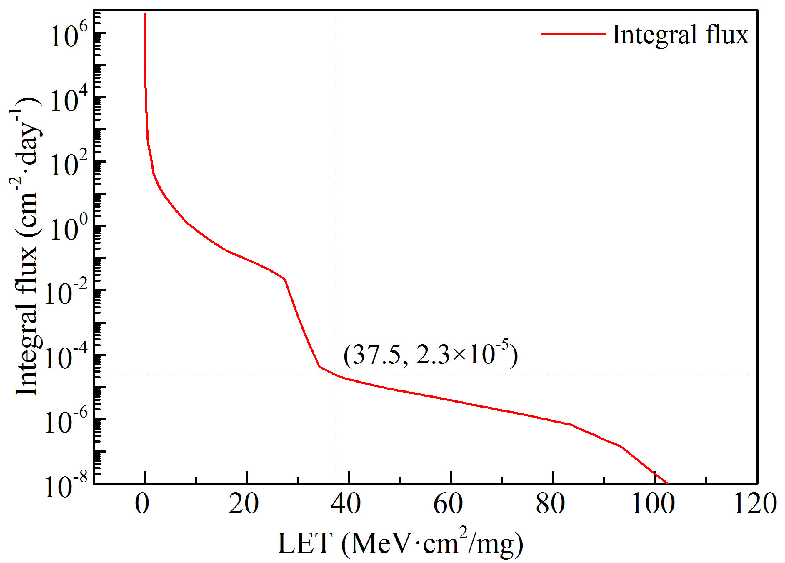}
\figcaption{\label{fig1}   Integral LET spectrum for silicon (calculated by CREME96). }
\end{center}

\section{Overview of VA160 and VATA160}

VA160 and VATA160, which are designed by IDEAS (Norway) and manufactured with the 0.35um CMOS technology processed on epitaxial silicon wafer, are able to cope with 32 channels of PMT dynode signals simultaneously \cite{lab8,lab9}. VATA160 shown in Fig.~\ref{fig2} is composed of VA part and TA part in a die. VA part, which is the exactly same as VA160, consists of 32 identical channels for charge measurement \cite{lab10}. The charge pulse from PMT dynode is fast amplified by a charge sensitive amplifier (CSA), shaped by a slow shaper (S), sampled by a sample-and-hold unit (S/H), and switched to a differential current output buffer (AOB) via an analogue multiplexer (AMUX). Besides, via the analogue de-multiplexer (ADEMUX), each channel could be calibrated by injecting the external calibration charge. TA part generates trigger information. The output of each CSA is directly coupled to the corresponding input of TA part. It is amplified via a programmable gain stage (G) and a fast shaping amplifier (FS), then discriminated by a low-sensitive discriminator (C). If the pulse height exceeds the common threshold, a trigger signal is generated. All channels share a common wire-ORed trigger output. There are a few digital circuits in the chips as well. Two 32-bit shift registers of VA part control AMUX and ADEMUX respectively. A 165-bit configuration register of TA part stores the operating settings.

\end{multicols}
%\ruleup
\begin{center}
\includegraphics[width=15cm]{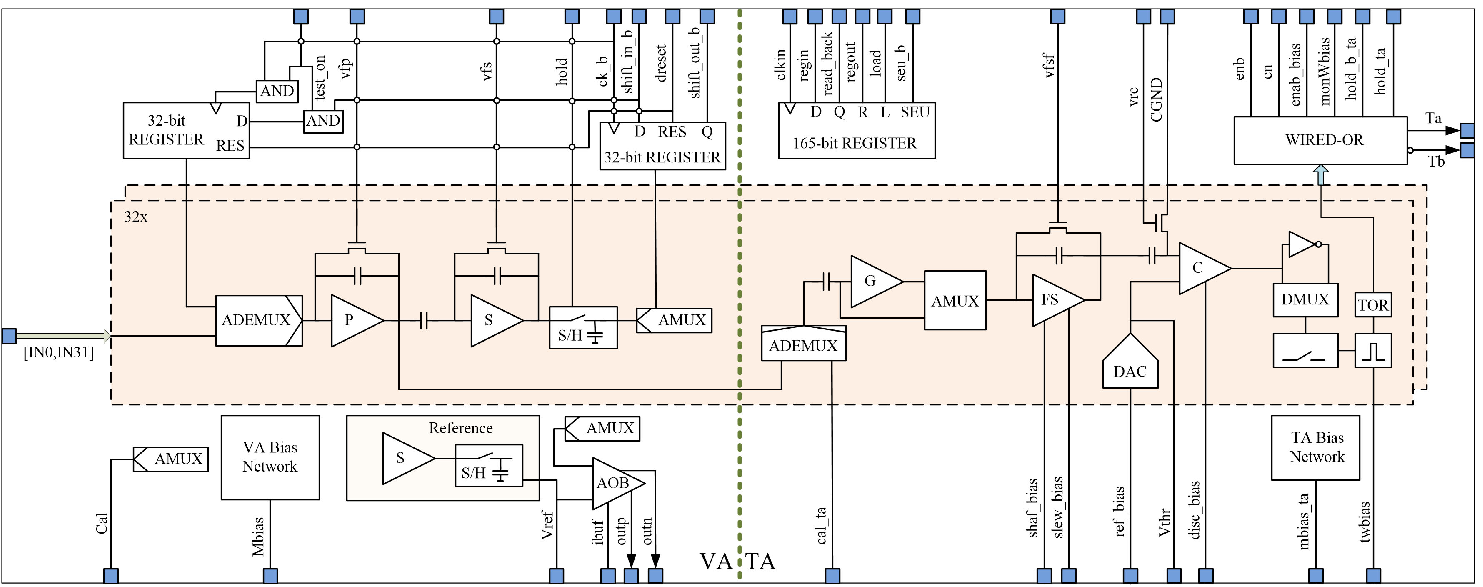}
\figcaption{\label{fig2}   Architecture of VATA160. }
\end{center}
%\ruledown

\begin{multicols}{2}

\section{Experimental approaches and results}

\subsection{Test setup}

In order to fit different irradiation facilities, the test setup shown in Fig.~\ref{fig3} is designed. Except device under test (DUT), DUT board includes a few passive components and connectors which are radiation-insensitive. VA160 and VATA160 have their own DUT boards respectively. DAQ board is much like a special IC tester that is designed for irradiation tests. It supplies power, sets work mode, monitors operating current, measures performance, and distinguishes abnormalities for DUT. An acquisition program on the remote computer controls DAQ board via RS-485 bus.
\\

\begin{center}
\includegraphics[width=8cm]{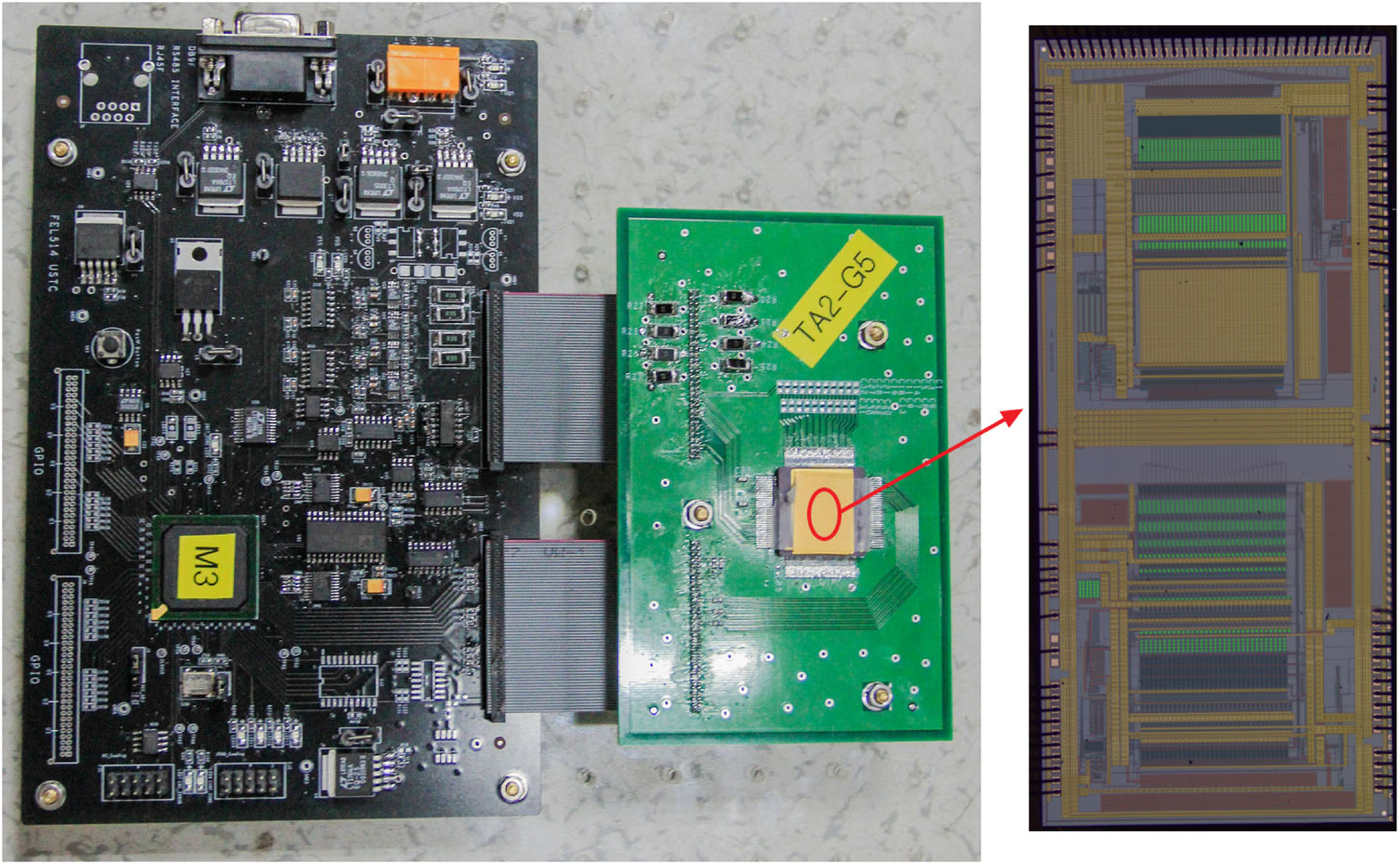}
\figcaption{\label{fig3}   Test setup and the die of VATA160. }
\end{center}

\subsection{Laser-induced SEE testing}
Laser-induced SEE testing is a much more cost-efficient way to rehearse heavy ion testing and verify the effectiveness of mitigation methods. Tests were performed at the Pulsed Laser Single Event Effects Facility (PLSEE) in National Space Science Center (Beijing). Pulsed laser with the wavelength of 1064 nm scanned the entire chip to explore latch-up phenomena \cite{lab11}.

Some interesting results were obtained. Firstly, as the spot diameter of the laser beam was less than 3 $\mu$m, observable latch-up sensitive areas were precisely located. Fig.~\ref{fig4} shows some of them. Secondly, the minimize laser energy triggering latch-up of VA part (or VA160) was greater than that of TA part, which inferred that threshold LET of VATA160 is lower. Thirdly, once a sensitive area was triggered, the current raised immediately to a steady value thereafter even though the same position was still under exposure. If the laser continued to irradiate other sensitive areas, the current raised step by step until the power shut off. However, pulsed laser is different in the mechanism of energy deposition from heavy ion, and forward incident laser could hardly reach the sensitive areas covered by the metal layers, heavy ion testing is indispensable \cite{lab12}.

\end{multicols}

\begin{center}
\includegraphics[width=15cm]{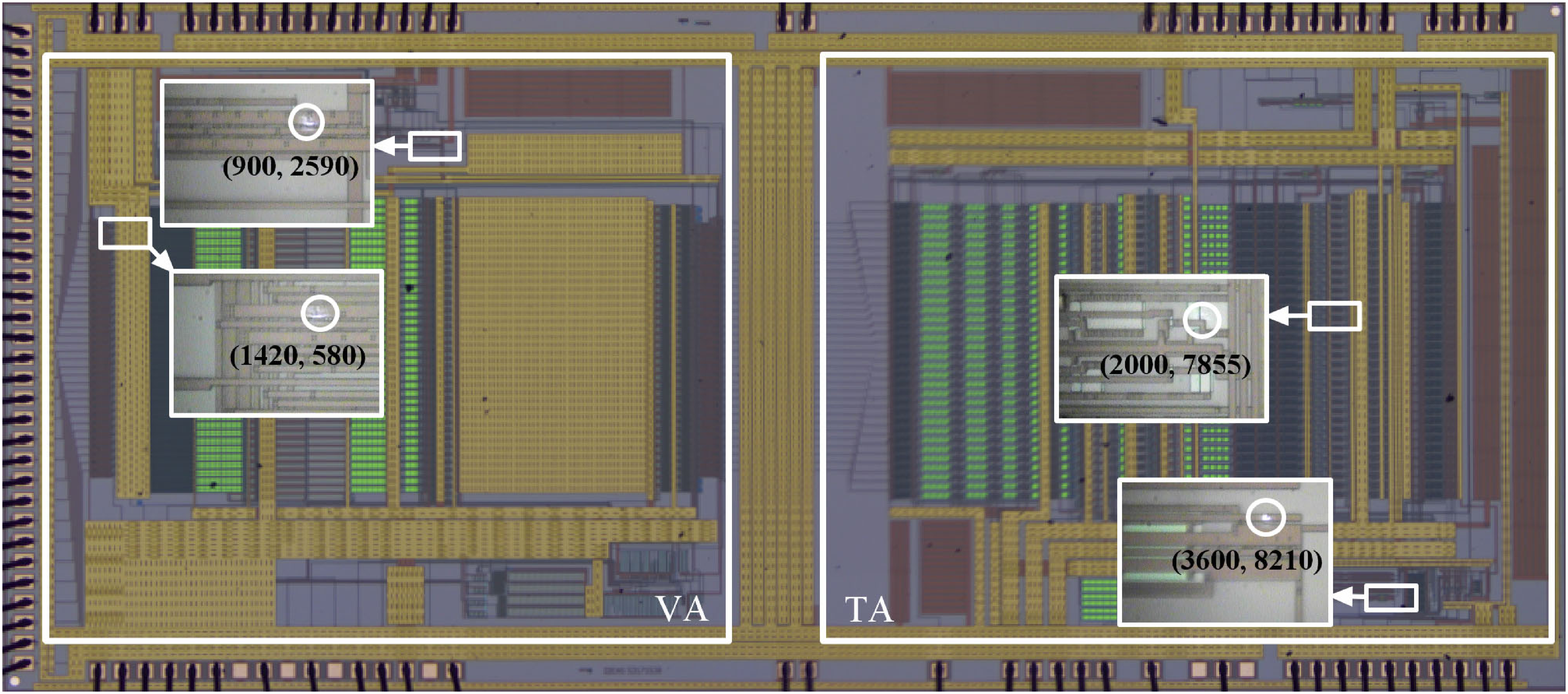}
\figcaption{\label{fig4}   Some SEL sensitive areas obtained from the laser testing. }
\end{center}

\begin{multicols}{2}

\subsection{Heavy ion-induced testing}
Heavy ion tests were performed at Heavy Ion Research Facility in Lanzhou (HIRFL, Lanzhou) and HI-13 tandem accelerator in China Institute of Atomic Energy (HI-13, Beijing). Irradiation tests performed at HIRFL were in air. It was convenient to observe all abnormal phenomena induced by ion strike. However, changing species or initial energy of ion is time-consuming, degraders were preferred to adjust the ion energy on the surface of die and thus specified LET values were obtained. Irradiation tests performed at HI--13 were in vacuum. DUT boards mounted inside the vacuum chamber were connected with DAQ boards through special adapters on the chamber. Since the number of adapters is limited, only power lines for DUTs were connected to investigate SEL. Five VA160 chips and three VATA160 chips with removal of the package lids were tested.

The test results of SEL are summarized in Table~\ref{tab1}. In order to get saturated cross section ($\sigma_{sat}$), threshold LET ($LET_{th}$), width factor (W) and shape factor (S), Weibull distributions are fitted for VA160 and VATA160 in Fig.~\ref{fig5} and Fig.~\ref{fig6} respectively. With these four parameters and the LET spectrum, SEL rates due to direct ionization are calculated to be: 7.5$\times$10$^{-5}$ /device/day for VA160 and 5.2$\times$10$^{-4}$ /device/day for VATA160.

\begin{center}
\includegraphics[width=8cm]{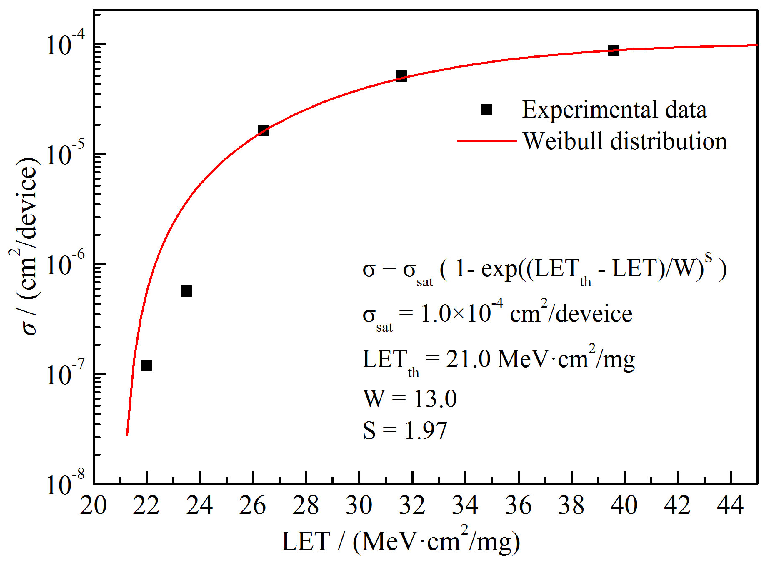}
\figcaption{\label{fig5}   SEL cross section of VA160 versus LET (calculated by CREME96). }
\end{center}

\end{multicols}

\begin{center}
\tabcaption{ \label{tab1}  Summary of SEL results.}
\footnotesize
\begin{tabular*}{170mm}{@{\extracolsep{\fill}}ccccccc}
\toprule Facility & DUT & Ion species & LET/(MeV$\cdot$cm$^2$/mg) & Fluence/(ions/cm$^2$) & Latchh-ups & Cross section/(cm$^2$/device)\\
\hline
HIRFL & VA160 & $^{84}$Kr & 20.6 & 2.23$\times$10$^{7}$ & 0 &  N/A \\
HIRFL & VA160 & $^{84}$Kr & 22.0 & 3.39$\times$10$^{7}$ & 4& 1.18$\times$10$^{-7}$ \\
HIRFL & VA160 & $^{84}$Kr & 23.5 & 5.40$\times$10$^{6}$ & 3 & 5.56$\times$10$^{-6}$\\
HIRFL & VA160 & $^{84}$Kr & 26.4 & 1.00$\times$10$^{5}$ & 16 & 1.60$\times$10$^{-5}$ \\
HIRFL & VA160 & $^{84}$Kr & 31.6 & 3.01$\times$10$^{5}$ & 15 & 4.98$\times$10$^{-4}$ \\
HIRFL & VA160 & $^{84}$Kr & 39.6 & 1.75$\times$10$^{5}$ & 15 & 8.57$\times$10$^{-4}$ \\
HI-13 & VATA160 & $^{13}$Al & 8.4 & 3.00$\times$10$^{7}$ & 0 & N/A \\
HI-13 & VATA160 & $^{32}$Cl & 13.1 & 2.75$\times$10$^{7}$ & 16 & 5.82$\times$10$^{-7}$ \\
HI-13 & VATA160 & $^{32}$Cl & 15.0 & 1.00$\times$10$^{7}$ & 24 & 2.40$\times$10$^{-6}$ \\
HI-13 & VATA160 & $^{22}$Ti & 21.8 & 9.22$\times$10$^{6}$ & 160 & 1.73$\times$10$^{-5}$ \\
HIRFL & VATA160 & $^{129}$Xe & 50.9 & 2.01$\times$10$^{5}$ & 82 & 4.08$\times$10$^{-4}$ \\
HIRFL & VATA160 & $^{129}$Xe & 64.5 & 1.67$\times$10$^{5}$ & 103 & 6.17$\times$10$^{-4}$ \\
\bottomrule
\end{tabular*}%
\end{center}

\begin{multicols}{2}

\begin{center}
\includegraphics[width=8cm]{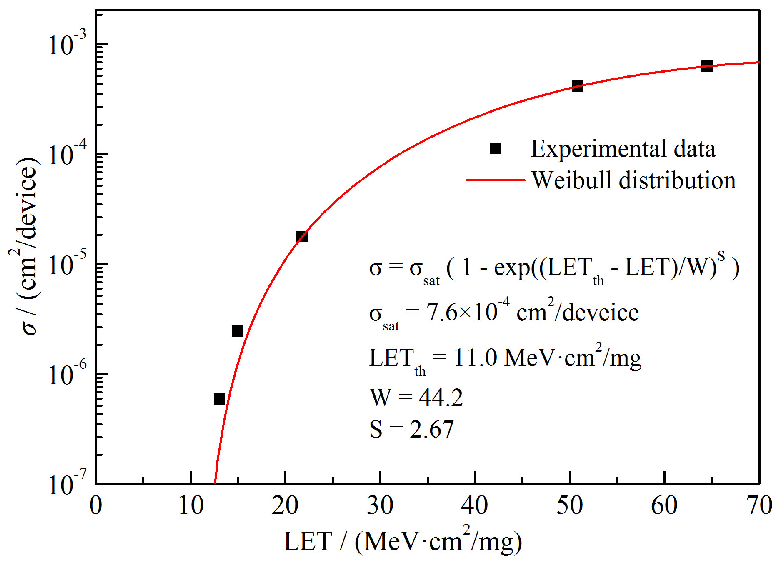}
\figcaption{\label{fig6}   SEL cross section of VATA160 versus LET (calculated by CREME96). }
\end{center}

The 165-bit configuration register of VATA160 was monitored when the testing was performed at HIRFL. No upset event was observed.

\section{Discussion of hardness assurance and mitigation methods}

\subsection{Proton SEL rate}
The maximum trapped proton energy in the Earth¡¯s radiation belts is around 400 MeV, thus the max LET of second particles produced by the inelastic interactions of protons with Si is about 13.0 MeV$\cdot$cm$^2$/mg \cite{lab13}. Since the SEL LET threshold of VATA160 is about 11.0 MeV$\cdot$cm$^2$/mg, the proton-induced latch-up should be considered. However, since there is no proper proton source available for us, an empirical PROFIT method that uses experimental data of heavy ion to predict the proton rate is adopted instead \cite{lab14}. Through calculation, the proton SEL rate is about 4.8$\times$10$^{-8}$ /device/day, which is far lower than the heavy ion SEL rate even multiplied by a tenfold calculation error. Hence, the impact of proton SEL is negligible.

\subsection{SEL protection Method}

Over the mission period of 3 years, 54 VA160 chips and 32 VATA160 chips applied in BGO calorimeter may suffer 4 SEL events and 18 SEL events respectively. Therefore, mitigation methods are strongly recommended. However, it is not practical to change the IC layout or the fabrication process because of cost and schedule. Preventing the chips from damage caused by SEL in the design of FEE is another possible way.

Though there are at most six VA160 or VATA160 chips mounted on a front-end board, the total fluctuant current is less than 10 mA no matter what mode the chips run in. And, the minimum latch-up current that has been observed is at least 25 mA larger than normal operating current. Therefore, a SEL protection circuit with fast respond shown in Fig.~\ref{fig7} is developed. A SEL event is identified once the current exceeds preset threshold, and then the comparison algorithm in FPGA disables LDO regulators immediately. The time from current over the threshold to power off is less than 100 $\mu$s, which minimizes the burden of power supply. During the power outage, all control signals from FPGA are set to ground level as well to avoid the emergence of sneak circuits. After latch-up is removed, LDO regulators are enabled again to power the chips and the work mode is subsequently restored to the status before power off.

\begin{center}
\includegraphics[width=7.5cm]{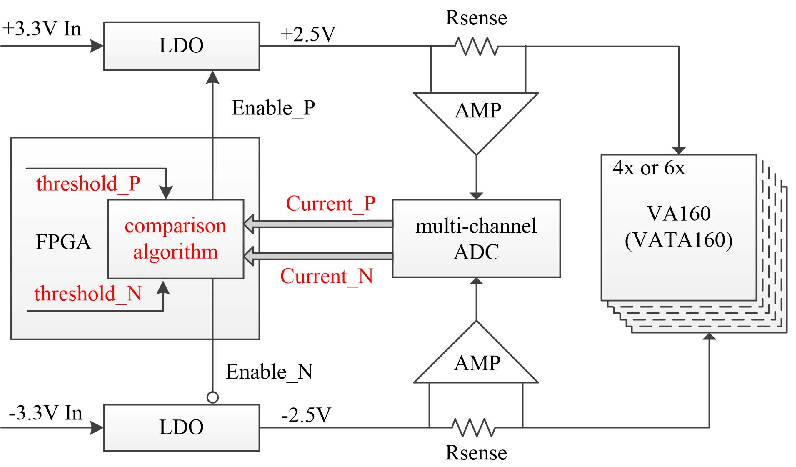}
\figcaption{\label{fig7}   Simplified schematic of the SEL protection circuit. }
\end{center}

A prototype of the front-end board with the protection circuit was verified by pulsed laser, as shown in Fig.~\ref{fig8}. Latch-ups on VA160 or VATA160 were identified and cleared thousands of times without function or performance damage, which sufficiently proved the effectiveness of the protection circuit.

\begin{center}
\includegraphics[width=8cm]{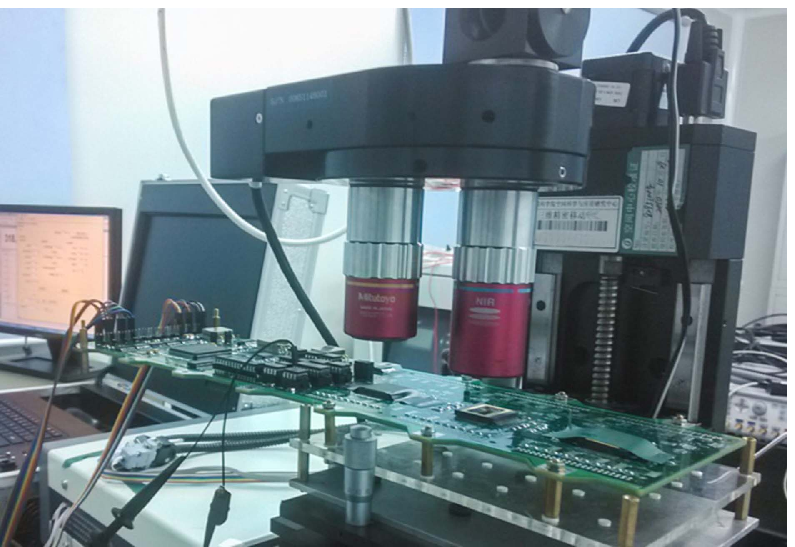}
\figcaption{\label{fig8}   SEL protection circuit in the front-end board was verified by pulsed laser. }
\end{center}

\subsection{SEU immune configuration registers}

The 165-bit configuration register in TA part, which maintains the settings for very long time in orbit, consists of a series of triple-redundancy flip-flops (TRFFs) implemented to avoid loss of information upon SEU events. A sketch showing the principle of a TRFF is in Fig.~\ref{fig9}. After all bits have been shifted into the DFFs of the serial shift register, the value of each DFF is loaded into the three parallel latches. The output of each TRFF is the logic value that the majority of the latches store.      \\ \begin{center}
\includegraphics[width=8cm]{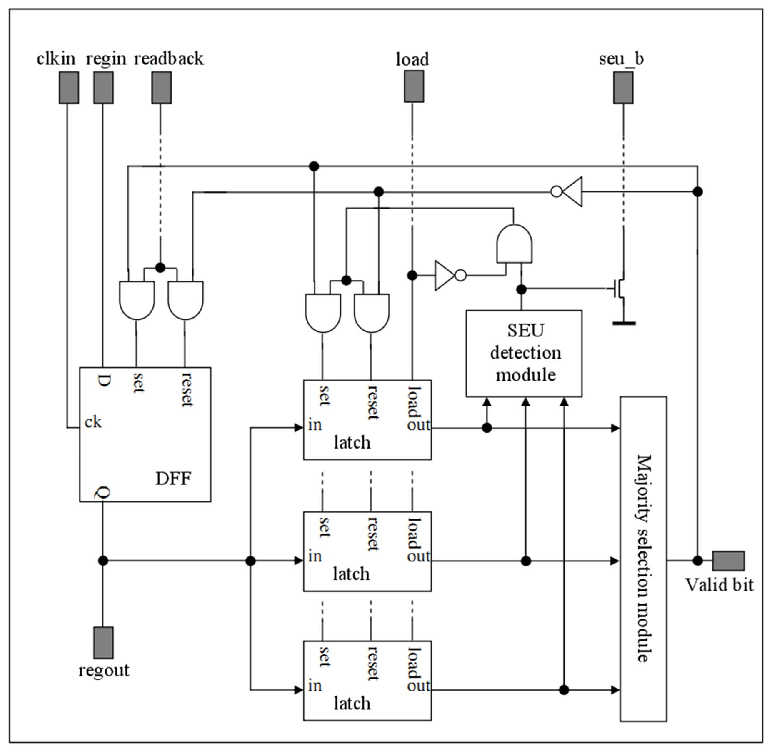}
\figcaption{\label{fig9}   The structure sketch of triple-redundancy flip-flops \cite{lab9}. }
\end{center} In case of an upset in a latch, the single event detection module automatically rewrites the three latches with the correct value and sends a flag signal (seu$\_$b) to the external system. A feedback mechanism is provided to load the output of TRFF into the DFF which could be readout through shift operation. During the heavy ion testing performed at HIRFL, the registers were readout and checked the moment a pulse appeared on the seu$\_$b pin. No error was found, which verified that the TRFF is immune to SEU.

\subsection{Optimized sequential design}

VA part (or VA160) has two identical 32-bit shift registers that execute the same timing operation shown in Fig.~\ref{fig10}. The registers are reset at the beginning and end of each acquisition, which prevents upset from accumulating and sustaining. As a result, each acquisition is independent. If a register is upset, the worst situation that could happen is that the current acquisition fails. Besides, the chance that the registers of two or more chips are upset simultaneously within an acquisition cycle (less than 1.0 ms) is too rare to happen. According to the physics simulation, temporary abnormality of a chip hardly affects the electromagnetic shower reconstruction in off-line data analysis. Hence, SEU that occurs on the shift registers is tolerable. Moreover, independent acquisition makes single event transient (SET) negligible though the analog circuit occupies the most die area of the ASICs \cite{lab15}.

\begin{center}
\includegraphics[width=8cm]{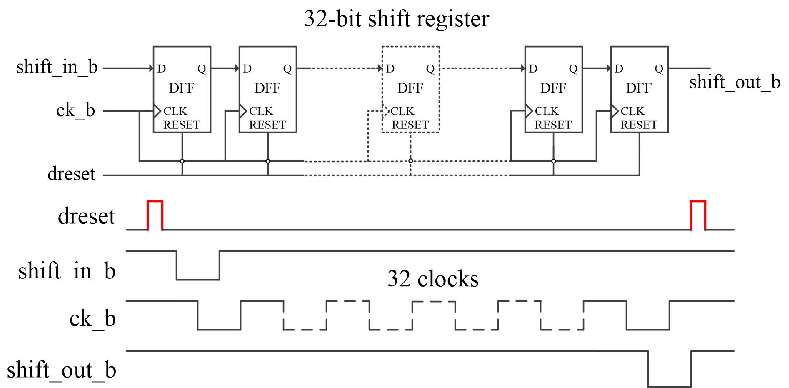}
\figcaption{\label{fig10}   Timing diagram of the 32-bit shift register. }
\end{center}

\section{Conclusion}

VA160 and VATA160 chips achieve all the requirements to implement the front-end electronics of BGO calorimeter except for unknown radiation tolerance, thus SEE tests with pulsed laser and heavy ion were performed. The results showed that the chips are SEL sensitive. The number of expected SEL events on orbit is not negligible since there are 54 VA160 chips and 32 VATA160 chips applied in the calorimeter for more than 3 years. Therefore, an effective SEL protection circuit with fast respond, which was sufficiently verified by pulsed laser, has been added into the FEE to avoid catastrophic damage caused by SEL. We also concluded that the 165-bit configuration register is immune to SEU, and periodic refreshing removes any possible soft errors caused by ion strike during the long-term acquisition in space. Benefiting from these mitigation methods, the flight model of BGO calorimeter achieves radiation hardness assurance.

\acknowledgments{We acknowledge the support and cooperation of the CAS Center for Excellence in Particle Physics (CCEPP), China Institute of Atomic Energy (CIAE), Institute of High Energy Physics(IHEP, CAS), and National Space Science Center (NSSC, CAS). $\cdots$.}

\end{multicols}

\vspace{-1mm}
\centerline{\rule{80mm}{0.1pt}}
\vspace{2mm}

\begin{multicols}{2}

\end{multicols}

\clearpage

%\end{CJK*}
\end{document}